\documentclass[fleqn,10pt]{wlscirep}
\usepackage[utf8]{inputenc}
\usepackage[T1]{fontenc}
\usepackage[utf8]{inputenc}
\usepackage{units}
\usepackage{amsmath}
\usepackage{physics}
\usepackage[utf8]{inputenc}
\usepackage[T1]{fontenc}\usepackage{physics}
\usepackage{graphicx}
\usepackage{float} 
\usepackage{amssymb}
\usepackage{graphicx}
\usepackage{bm}
\usepackage{multirow,microtype,color,relsize,ulem}

\title{Self-trapping and skin solitons in two-dimensional non-Hermitian lattices}

\author[1,2]{Emmanouil T. Kokkinakis}
\author[1,2]{Ioannis Komis}
\author[1,2,*]{Konstantinos G. Makris}

\affil[1]{Department of Physics, University of Crete, 70013 Heraklion, Greece}
\affil[2]{Institute of Electronic Structure and Laser (IESL), FORTH, 71110 Heraklion, Greece}

\affil[*]{makris@physics.uoc.gr}

\begin{abstract}

Two-dimensional non-Hermitian photonic lattices with asymmetric couplings offer rich possibilities for controlling wave localization, through the emergence of the non-Hermitian skin effect at lattice corners or sides. Incorporating optical nonlinearity fundamentally alters these boundary-localization characteristics. Here we show that in a two-dimensional Hatano–Nelson lattice with Kerr nonlinearity, the interplay between self-trapping and directional propagation leads to position-dependent amplitude thresholds. Single-site excitations having above a critical amplitude become confined to their initial position, with lower thresholds near the position where the linear eigenmodes are localized and higher thresholds within the lattice's bulk. Additionally, we study the differences of this dynamical interplay, for wider initial excitations, between the focusing and defocusing Kerr-nonlinearity regimes. Lastly, we identify skin soliton solutions in a variety of two-dimensional lattice geometries featuring coupling asymmetry. 
\end{abstract}
\begin{document}

\flushbottom
\maketitle
%
%

\section*{Introduction}

In recent years, photonics has emerged as a versatile platform for exploring non-Hermitian physics, enabling the precise engineering of gain and loss distributions in optical systems \cite{el_ganainy_2018}. Unlike quantum or condensed-matter systems, where the experimental control of non-Hermiticity remains challenging, photonic systems offer a practical and accessible setup \cite{ruter_2010}. This has fueled rapid progress in the study of non-Hermitian wave dynamics and the demonstration of novel phenomena, such as exceptional points \cite{EP2, EP3}, parity-time $(\mathcal{PT})$ symmetry \cite{PT1, PT2, PT3, PT4, PT5, PT6, PT7, PT8, PT9, komis_2024}, and the Non-Hermitian Skin Effect (NHSE) \cite{NHSE1, NHSE2, NHSE3, NHSE4}, among other \cite{NHJumps_1, NHJumps_2, Molignini2023, Delgado2025, Hu2025}, all of which are now of crucial importance and relevance to the fields of topological and integrated photonics.

Among the plethora of systems studied in non-Hermitian physics, the Hatano–Nelson (HN) model offers a prototypical lattice where non-Hermiticity is realized through asymmetric nearest-neighbor couplings \cite{HN1, HN2}. An intriguing feature of this model is the NHSE, i.e., the exponential localization of eigenmodes at the boundary of the system under open boundary conditions (OBC). Initially proposed to describe localization–delocalization transitions in solid-state systems, the HN model remained for years a purely theoretical concept in mathematical physics \cite{HN3}. However, recent advances in optics have enabled its experimental realization \cite{HN_Exp1, HN_Exp2}, opening new possibilities for practical applications in photonic cavities and lasers \cite{HN_Exp3, HN_Exp4, HN_Exp5, HN_Exp6, topo10} as well as ultracold atoms \cite{Fermi}.
The extension of the NHSE to two-dimensional (2D) systems reveals even richer localization features that depend on the direction and strength of coupling asymmetry \cite{2D_1, 2D_2, 2D_3, 2D_4, 2D_5}. When the coupling asymmetry is along a single direction, the NHSE manifests as side-localized modes; however, when asymmetric couplings are present in both directions, the eigenmodes exhibit significant spatial asymmetry as they become localized at specific lattice corners. Notably, most studies related to higher-dimensional lattices have focused on the topological nature of the NHSE, exploring bulk–boundary correspondence and redefining topological invariants \cite{Topo1, Topo2, Topo3, Topo4, Topo5, Topo6, Topo7}, while much less attention has been devoted to transport features \cite{2D_4}.

On the other hand, nonlinearity, an inherent property of optical systems, gives rise to a wide range of intriguing phenomena, including self-trapping, modulational instability, and soliton formation \cite{NLH1}. When combined with non-Hermiticity, the nonlinear phenomena are significantly modified, leading to novel effects and the emergence of localized states \cite{NLNH1, PRE, Science}. In one-dimensional (1D) HN systems, nonlinearity has primarily been explored in the context of topological edge states and single-mode lasing \cite{NL1, NL2, NL3, NL4}. Recent studies have also examined the effect of Kerr nonlinearity in 1D HN lattices, focusing on soliton formation \cite{SL, SL_Exp, manda_2024, longhi_2025, yuce_2025, Ghaemi2024}. Notably, it was recently shown, both theoretically and experimentally, that the interplay between Kerr nonlinearity and asymmetric couplings can lead to the formation of nonlinear skin solitons \cite{SL, SL_Exp, Li2025}, characterized by position-dependent power thresholds and highly asymmetric spatial profiles. Nevertheless, the interplay of nonlinearity and non-Hermiticity in higher-dimensional systems, such as 2D HN lattices, remains largely unexplored.

In this work, we study the dynamics in 2D HN lattices with Kerr nonlinearity, focusing on the antagonism between self-trapping and propagation due to asymmetric couplings. In the extensively studied Hermitian case, where the evolution is described by the discrete nonlinear Schrödinger equation (DNLSE), self-trapping of a localized excitation to its initial position occurs when the excitation's amplitude exceeds a critical threshold, thus suppressing discrete diffraction. This phenomenon has been widely explored in both 1D and 2D systems within the framework of lattice solitons \cite{Spatial_Segev, 2D_Segev, Surface_Solitons, Solitons_Optics, PT_Solitons1, PT_Solitons2, Power_Thre}. In particular, here we examine how the degree of non-Hermiticity and the location of the initial single-site excitation influence the amplitude thresholds for self-trapping. Our study reveals that these thresholds are highly position-dependent, with relatively low thresholds near the lattice corners toward which the couplings are stronger, and significantly higher thresholds—or even impossible self-trapping—in the bulk of the system under strong non-Hermiticity. {Furthermore, we consider wider Gaussian initial excitations and investigate how the interplay between Kerr nonlinearity and non-Hermiticity depends on whether the nonlinearity is focusing or defocusing.} Additionally, we identify two-dimensional skin soliton solutions and characterize their properties through their corresponding power–eigenvalue diagrams. Our results show that these skin solitons exhibit spatial asymmetry, and their power thresholds increase monotonically as the coupling asymmetry becomes stronger. Finally, we report the existence of skin solitons in a variety of more complex two-dimensional nonlinear non-Hermitian lattices, beyond the 2D HN lattice.

\section*{Results}

\subsection*{Two-dimensional nonlinear Hatano-Nelson lattice}

\subsubsection*{Coupled-mode equations and conservation laws}
 
We begin our study with a two-dimensional generalized nonlinear Hatano-Nelson lattice  (2D NLHN) consisting of \( N \times N \) coupled waveguides, indexed by \( (n_{x}, n_{y}) \in \{1, 2, \dots, N\} \), and characterized by asymmetric couplings. The discrete paraxial equation describing the wave evolution in this lattice, in normalized units, is  
\begin{equation}
\label{kerr}
\begin{split}
i\frac{\partial \psi_{n_{x},n_{y}}}{\partial z} + e^{-h_{x}}\psi_{n_{x}+1,n_{y}} + e^{h_{x}}\psi_{n_{x}-1,n_{y}} +  e^{-h_{y}}\psi_{n_{x},n_{y}+1}  
+ e^{h_{y}}\psi_{n_{x},n_{y}-1} + g|\psi_{n_{x},n_{y}}|^2 \psi_{n_{x},n_{y}} = 0
\end{split}
\end{equation}
where \( h_{x}, h_{y}\in \mathbb{R}^{+} \) are the non-Hermiticity parameters, in the horizontal (\(n_{x}\)) and vertical (\(n_{y}\)) directions, respectively. The quantity \( \psi_{n_{x},n_{y}}(z) \) denotes the complex amplitude of the electric field's envelope at site \( (n_{x},n_{y}) \) at propagation distance \( z \). The parameter $g$ determines whether the lattice is linear ($g=0$) or exhibits Kerr-type nonlinearity ({$|g|=1$}). Throughout this work, we consistently consider OBC in both lattice directions, i.e., $\psi_{0,n_y}=\psi_{N+1,n_y}=\psi_{n_x,0}=\psi_{n_x,N+1}=0$. \begin{figure}[H]
    \centering
    \includegraphics[width=0.6\textwidth]{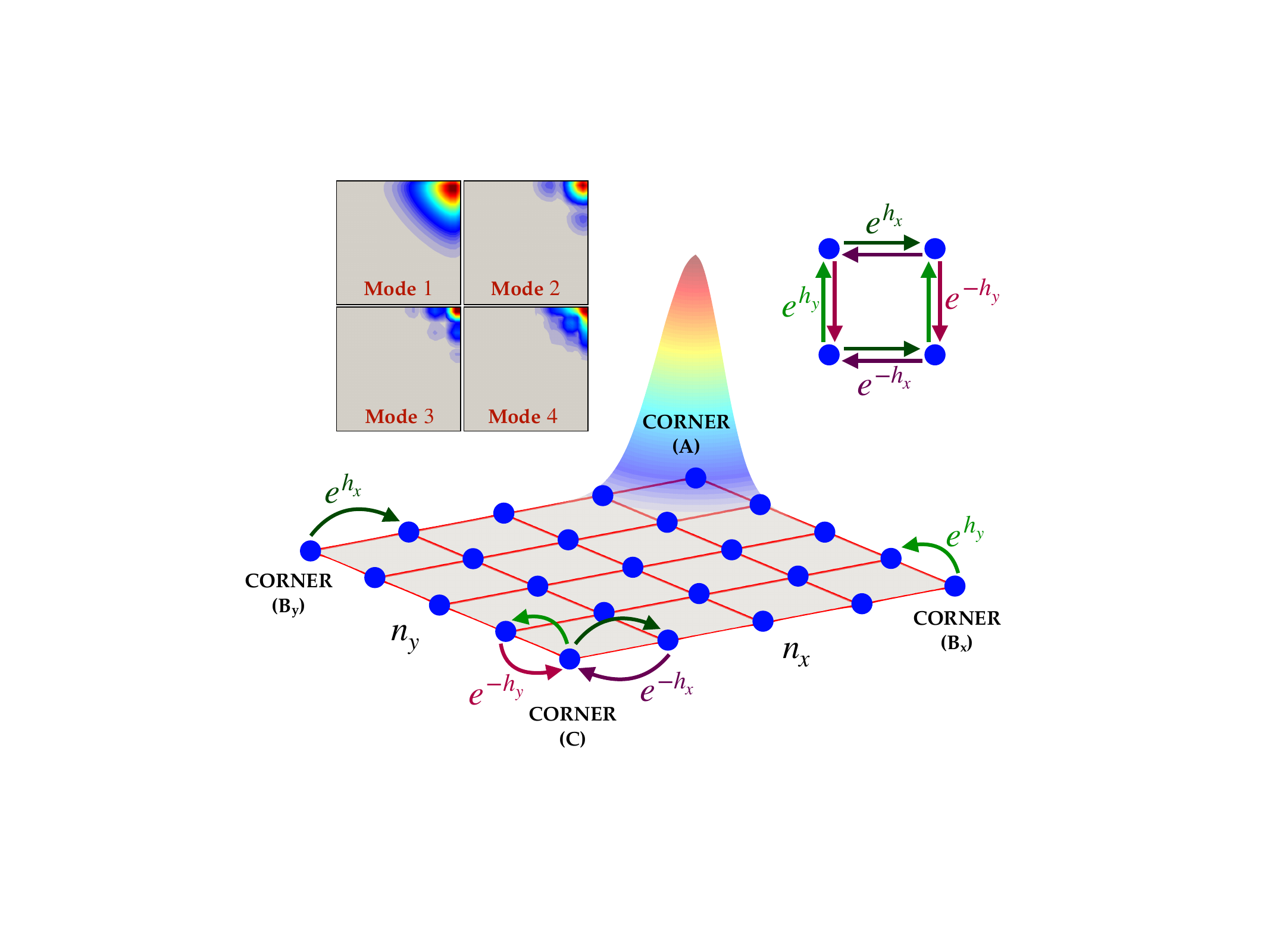}
    \caption{Schematic representation of a two-dimensional Hatano-Nelson lattice. The arrows represent the asymmetric couplings: $e^{h_x}$ and $e^{-h_x}$ along the $n_x$-direction, and $e^{h_y}$ and $e^{-h_y}$ along the $n_y$-direction. The figure identifies the four geometrically distinct corners of the lattice, labeled as (A), (B$_x$), (B$_y$), and (C). The spatial distribution of four indicative linear eigenmodes, for a case with $h_x=h_y$, is included as inset.}
    \label{fig:Schematic}
\end{figure}

The geometry of such a square lattice, schematically depicted in Fig.~\ref{fig:Schematic}, implies the existence of four geometrically distinct corners. These corners are labeled as follows: corner (A), located in the direction of stronger coupling along both $n_x$ and $n_y$, where the linear skin modes are localized; corner (C), located in the direction of weaker coupling along both $n_x$ and $n_y$; and corners (B$_x$)/(B$_y$), located in the directions of stronger coupling along $n_x$/$n_y$ but weaker coupling along $n_y$/$n_x$. In the special case where $h_x = h_y$, corners B$_x$ and B$_y$ are equivalent by symmetry, and we refer to them in such case as corners (B). We consistently follow this notation throughout this work. 

Although the 2D NLHN is non-Hermitian and thus corresponds to non-conservative dynamics, it can be mapped via the imaginary gauge transformation \(\alpha_{n_{x},n_{y}} \equiv \psi_{n_{x},n_{y}} e^{-(h_{x}n_{x}+h_{y}n_{y})}\), to a Hermitian tight-binding model, which, in the nonlinear case ({$|g|=1$}), is characterized by site-dependent nonlinearity. As we will see, this allow us to derive conservation laws for the system under discussion.

Specifically, the evolution of the complex amplitude $\alpha_{n_{x},n_{y}}(z)$ in such a lattice is governed by the discrete nonlinear Schrödinger (DNLSE)-type equation:  

\begin{equation}
\label{periodic_2d_NL}
\begin{split}
i\frac{\partial \alpha_{n_{x},n_{y}}}{\partial z} + \alpha_{n_{x}+1,n_{y}} + \alpha_{n_{x}-1,n_{y}} +  \alpha_{n_{x},n_{y}+1} 
+ \alpha_{n_{x},n_{y}-1} + ge^{2(h_xn_{x}+h_yn_{y})} |\alpha_{n_{x},n_{y}}|^2 \alpha_{n_{x},n_{y}} = 0.
\end{split}
\end{equation}
As detailed in the Supplemental Material (SM), the quantity
\begin{equation}
    \mathcal{P}(z) \equiv \sum_{n_{x}=1}^{N}\sum_{n_{y}=1}^{N} |\alpha_{n_{x},n_{y}}(z)|^2
\end{equation}
which represents the optical power and is known to be conserved in the standard DNLSE, remains conserved even under this particular site-dependent nonlinearity. This contrasts with the Hatano-Nelson system of Eq.~(\ref{kerr}), in which the optical power
$\mathcal{P}_{\text{HN}}(z) = \sum_{n_{x}=1}^{N}\sum_{n_{y}=1}^{N} |\psi_{n_{x},n_{y}}(z)|^2$
is not conserved along propagation. Nonetheless, the connection between the systems described by Eq.~(\ref{kerr}) and Eq.~(\ref{periodic_2d_NL}) through the transformation  
$\alpha_{n_{x},n_{y}} \equiv \psi_{n_{x},n_{y}} e^{-(h_{x}n_{x}+h_{y}n_{y})},
$
along with the power conservation of the latter system, implies that in the 2D NLHN lattice [Eq.~(\ref{kerr})], the quantity 
\begin{equation}
    \label{power}
    \tilde{\mathcal{P}} = \sum_{n_{x}=1}^{N}\sum_{n_{y}=1}^{N} |\psi_{n_{x},n_{y}}|^2 e^{-2(h_{x}n_{x}+h_{y}n_{y})}
\end{equation}
which we refer to as the \textit{pseudopower}, is constant during the dynamics. 
Moreover, another invariant quantity of the system described by Eq.(\ref{periodic_2d_NL}) is the Hamiltonian,
\begin{equation}
\label{hamiltonian_herm}
\begin{split}
    \mathcal{H} \equiv & \sum_{n_{x}=1}^{N-1} \sum_{n_{y}=1}^{N-1} \big( 
    \alpha_{n_x,n_y}\alpha^{*}_{n_{x}+1, n_y}
    + \alpha_{n_x,n_y}\alpha^{*}_{n_{x}, n_y+1}  
    + \text{c.c.} \big) + \sum_{n_{x}=1}^{N} \sum_{n_{y}=1}^{N} ge^{2(h_xn_{x}+h_yn_{y})}\frac{|\alpha_{n_x,n_y}|^4}{2},
\end{split}
\end{equation}
as shown in the SM.  Consequently, a second conserved quantity for the 2D NLHN lattice is the \textit{pseudohamiltonian}
\begin{equation}
\label{hamiltonian}
\begin{split}
    \tilde{\mathcal{H}} = & \sum_{n_{x}=1}^{N-1} \sum_{n_{y}=1}^{N-1} 
    e^{-2(h_xn_x + h_yn_y)} \Big( 2e^{-h_x} \, \text{Re}(\psi_{n_x,n_y}^{*} \psi_{n_x + 1, n_y}) + 2e^{-h_y} \, \text{Re}(\psi_{n_x,n_y}^{*} \psi_{n_x, n_y + 1}) 
       \Big)  + \sum_{n_{x}=1}^{N-1} \sum_{n_{y}=1}^{N-1}g e^{-2(h_xn_x + h_yn_y)}  \frac{|\psi_{n_x,n_y}|^4}{2} 
\end{split}
\end{equation}
Within the context of our discussion, the existence of these two conservation laws provides a direct means of validating the accuracy of our numerical results, which will be presented in the following subsections. This is very important, given the high degree of non-normality and numerical instability due to coupling asymmetry, that are inherent in Hatano-Nelson lattices \cite{Feng2025}. 
\subsubsection*{Impact of nonlinearity on propagation dynamics}
In this paragraph, we investigate how the Kerr nonlinearity affects the dynamics in the 2D NLHN lattice. Unless stated differently, in the numerical results presented in this study, we consider a lattice consisting of $N\times N = 25\times25$ waveguides and equal non-Hermiticity parameters for the two directions, i.e., $h_x=h_y\equiv h$. Complementary results for the case of $h_x\neq0$ and $h_y=0$ are discussed in the SM. In particular, we will consider the scenario of a localized, single-site initial excitation. {As analytically proven in the SM, for such an initial condition, the dynamics of the magnitude of the wavefunction $|\psi_{n_x,n_y}(z)|$ remain identical regardless of whether the Kerr nonlinearity is focusing ($g=1$) or defocusing ($g=-1$). Relevant examples of propagation dynamics with initial excitations occupying multiple channels are discussed in a separate subsequent subsection}.
\begin{figure}[H]
    \centering
    \includegraphics[width=1\textwidth]{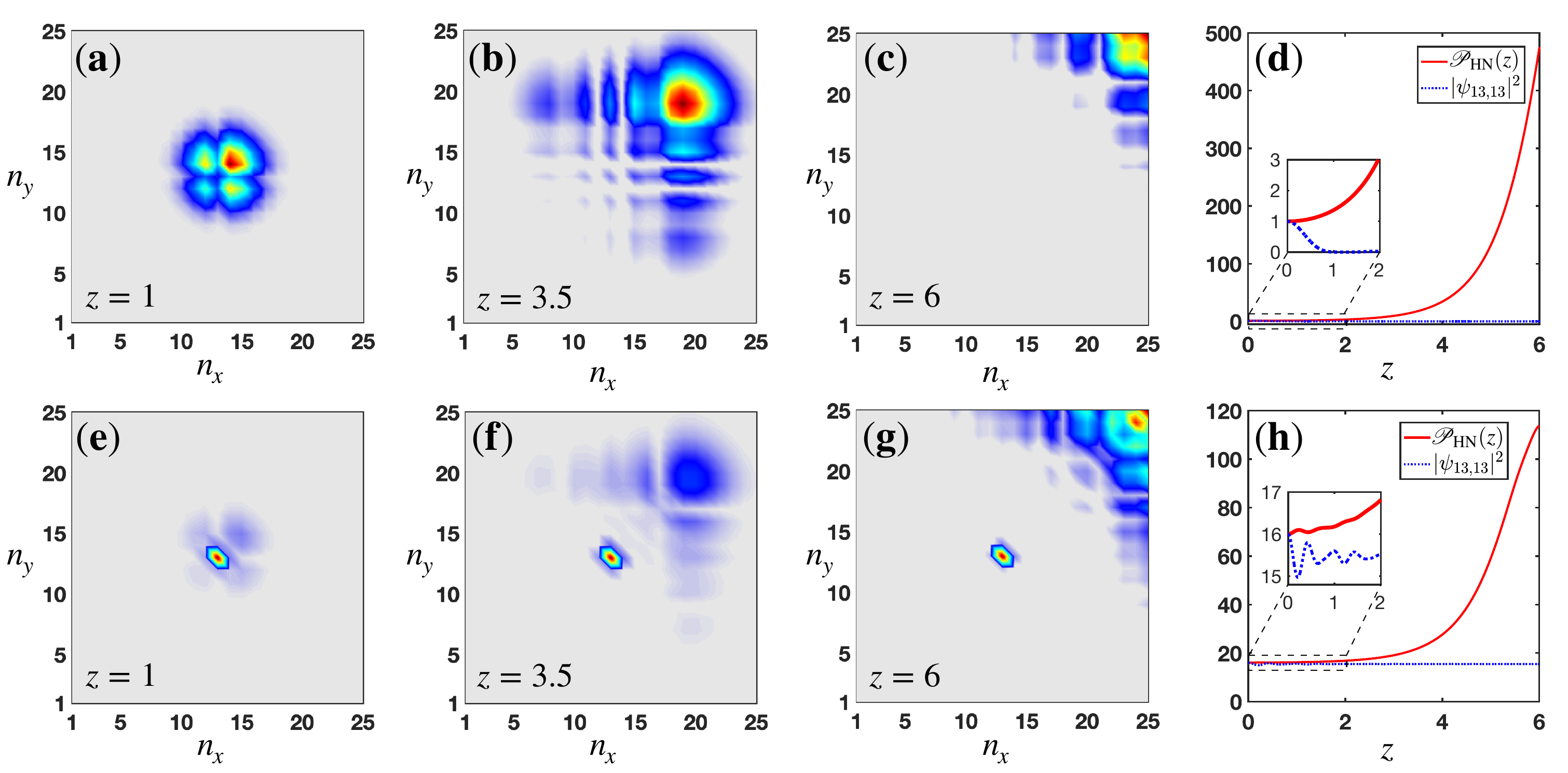}
    \caption{Propagation in a 2D HN lattice for the linear (top row) and Kerr-nonlinear (bottom row) cases, with non-Hermiticity parameter $h = 0.2$ and initial condition \( \psi_{n_x,n_y}(z=0) = A\delta_{n_x,13} \delta_{n_y,13} \). (a)–(c) Normalized complex amplitude $|\psi_{n_x, n_y}(z)|$ for the linear case at propagation distances (a) $z = 1$, (b) $z = 3.5$, and (c) $z = 6$. (e)–(g) Normalized complex amplitude $|\psi_{n_x, n_y}(z)|$ for the nonlinear case with input amplitude $A=4$, shown at the same propagation distances as in the top row. {Panels (d) and (h) show the evolution of the optical power $\mathcal{P}_{\text{HN}}$ (red lines) and the optical intensity at the site of initial excitation $|\psi_{13,13}|^2$ (blue lines) for the linear and nonlinear cases, respectively.}}
    \label{dynamics}
\end{figure}
First, as shown in the top row of Fig.~\ref{dynamics}, in the absence of Kerr nonlinearity ($g=0$), the NHSE dictates the dynamics. Specifically, for a non-Hermiticity parameter \(h = 0.2\) and an initial condition $\psi_{n_x,n_y}(z = 0) = A\,\delta_{n_x,13}\delta_{n_y,13}$, where \(A\) is the excitation amplitude, the wavefunction propagates toward corner (A) of the lattice, as physically expected. {As clearly demonstrated in Fig.~2(d), while the wavefunction shifts toward corner (A), the total optical power $\mathcal{P}_{\text{HN}}= \sum_{n_{x}=1}^{N}\sum_{n_{y}=1}^{N} |\psi_{n_{x},n_{y}}(z)|^2$ increases, consistent with the conservation of the pseudopower $\tilde{\mathcal{P}} = \sum_{n_{x}=1}^{N}\sum_{n_{y}=1}^{N} |\psi_{n_{x},n_{y}}(z)|^2 e^{-2(h_{x}n_{x}+h_{y}n_{y})}$ , since the wavefunction is on average distributed on lattice sites with larger $n_x$, $n_y$. Additionally, the optical intensity at the initially excited site, $|\psi_{13,13}|^2$, vanishes for $z>1$}. Of course, in the linear regime, diffraction is independent of the excitation amplitude \(A\), unlike the nonlinear case {($|g|=1$)}, where its value plays a crucial role. In particular, as illustrated in the bottom row of Fig.~\ref{dynamics}, for the same initial condition, with amplitude \(A=4\), the wavefunction remains self-trapped at the initial-excitation site for short propagation distances (\(z < 2\)). For $z \gtrsim 2$, it splits into two distinct parts: one remains localized at the initial position, while the other propagates toward corner (A). {This occurs because, in the Kerr nonlinear case, the optical intensity $|\psi_{13,13}|^2$ exhibits oscillations for $z < 2$ [Fig.~2(h)] and eventually stabilizes at a slightly reduced value compared to its initial one. The separated portion of optical power propagates toward the preferential corner and becomes amplified due to the conservation of pseudopower $\mathcal{\tilde{P}}$, as in the linear regime. This process results in wavefunction splitting: part of the initial power, slightly less than the original, remains localized near the initial-excitation site, while the rest is amplified as it propagates toward the preferential corner}. This behavior clearly indicates an antagonism between self-trapping caused by nonlinearity, and the NHSE, induced by the asymmetric couplings.

In order to better understand this interplay, we study the evolution of the wavefunction's mean positions and uncertainties along the \( n_x \) and \( n_y \) directions. These are defined by the following general expressions:

\begin{equation}
\langle n_{x/y} \rangle \equiv \frac{\sum_{n_x=1}^{N} \sum_{n_y=1}^{N} n_{x/y} |\psi_{n_x, n_y}|^2}{\sum_{n_x=1}^{N} \sum_{n_y=1}^{N} |\psi_{n_x, n_y}|^2},
\end{equation}

\begin{equation}
\Delta n_{x/y} \equiv \sqrt{\langle n_{x/y}^2 \rangle - \langle n_{x/y} \rangle^2},
\end{equation}
where
\begin{equation}
\langle n_{x/y}^2 \rangle \equiv \frac{\sum_{n_x=1}^{N} \sum_{n_y=1}^{N} n_{x/y}^2 |\psi_{n_x, n_y}|^2}{\sum_{n_x=1}^{N} \sum_{n_y=1}^{N} |\psi_{n_x, n_y}|^2}.
\end{equation}
In the above expressions, the notation \(n_{x/y}\) denotes either \(n_x\) or \(n_y\). 

\begin{figure}[H]
    \centering
    \includegraphics[width=0.7\textwidth]{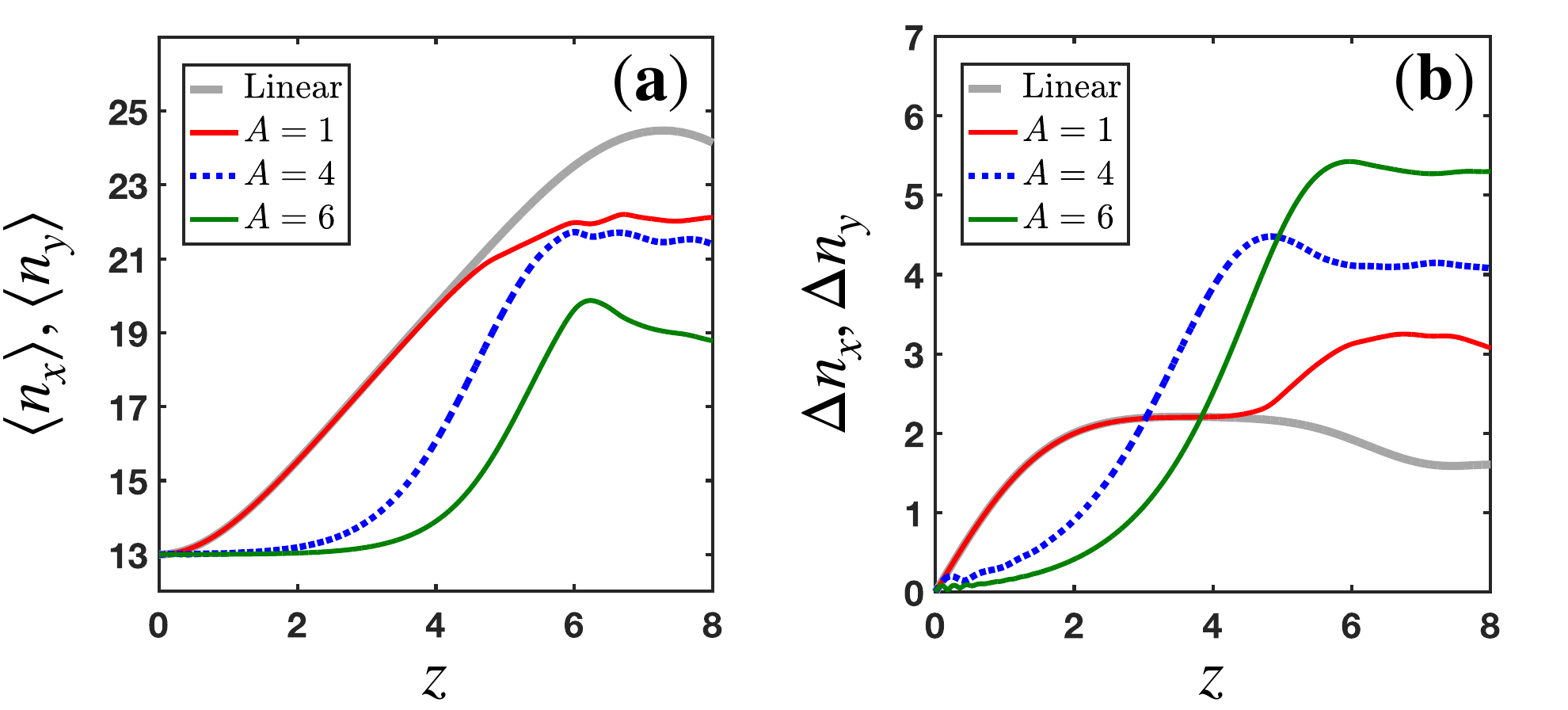}
    \caption{Evolution of mean positions and uncertainties in a 2D HN lattice with non-Hermiticity parameter $h= 0.2$, for the initial condition \( \psi_{n_x,n_y}(z=0) = A\delta_{n_x,13} \delta_{n_y,13} \). (a) Mean positions \( \langle n_x \rangle \) and \( \langle n_y \rangle \)  and (b) position uncertainties \( \Delta n_x \) and \( \Delta n_y \) for the linear case (gray line) and nonlinear cases with excitation amplitudes \( A=1 \) (red line), \( A=4 \) (blue line), and \( A=6 \) (green line).}
    \label{mean_std}
\end{figure}

Our corresponding results, for the same configuration as in Fig.2,  are presented in Fig.3. In particular, in Fig.~3(a), we show that, in the linear regime, the wavefunction's mean positions, \( \langle n_x \rangle = \langle n_y \rangle \), increase almost linearly with the propagation distance \( z \), after an initial acceleration~\cite{Topo6}, and finally the wavepacket reaches the corner (A). To the contrary, when Kerr nonlinearity is introduced, propagation is significantly hindered. Specifically, as the excitation amplitude \(A\) increases, the mean positions \( \langle n_x \rangle = \langle n_y \rangle \) have a progressively delayed increase and their value never reaches the corresponding for the linear regime, indicating a higher tendency for self-trapping. Moreover, the nonlinear regime is characterized by increased position uncertainties \( \Delta n_x = \Delta n_y \), as depicted in Fig.~\ref{mean_std}(b). This increase reflects the previously discussed splitting of the wavefunction into two parts (bottom row of Fig.~\ref{dynamics}). Such behavior emerges due to the interplay between Kerr nonlinearity and the NHSE, which will be studied more systematically in the following subsection. 

\subsection*{Antagonism between skin effect and self-trapping}

As it was discussed in the previous section, Kerr-nonlinearity and the skin effect have opposite tendencies regarding localization of an initially localized wavepacket, which leads to a dynamic interplay.  Therefore, our goal in this section is to demonstrate, through pertinent examples, that the dynamics in a 2D NLHN lattice, as described by Eq.~(\ref{kerr}) with {$|g|=1$}, depend highly on the position of the initial single-site excitation, due to the system's non-Hermiticity. Specifically, we will consider the single-site initial condition:
\begin{equation}
    \psi_{n_{x},n_{y}}(z=0) = A \delta_{n_{x},n_{x_{0}}} \delta_{n_{y},n_{y_{0}}}
\end{equation}
for different cases of $n_{x_0}$ and $n_{y_0}$ and for various values of the excitation amplitude \( A \). 

\begin{figure}[H]
    \centering
    \includegraphics[width=1\textwidth]{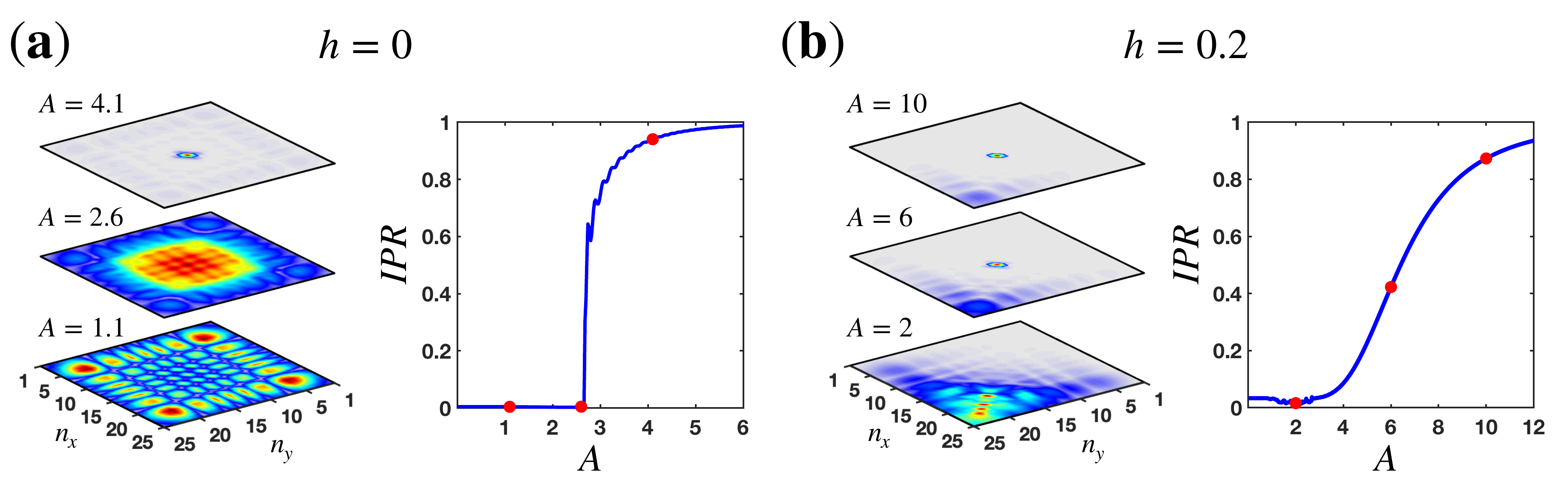}
    \caption{Comparison of the self-trapping tendency between (a) a Hermitian ($h=0$) and (b) a non-Hermitian ($h=0.2$) lattice, for single-site excitation at the center, i.e., \( \psi_{n_{x},n_{y}}(z=0) = A\,\delta_{n_{x},13} \delta_{n_{y},13} \). The normalized amplitudes \( |\psi_{n_{x},n_{y}}(z=5)| \) are plotted for pertinent values of \( A \) (left panels) and the $IPR$ for $z=5$ is shown as a function of \( A \) (right panels). Red dots correspond to the $A$ values used in the left panels.}
    \label{fig_4}
\end{figure}

\begin{figure}[H]
    \centering
    \includegraphics[width=0.94\textwidth]{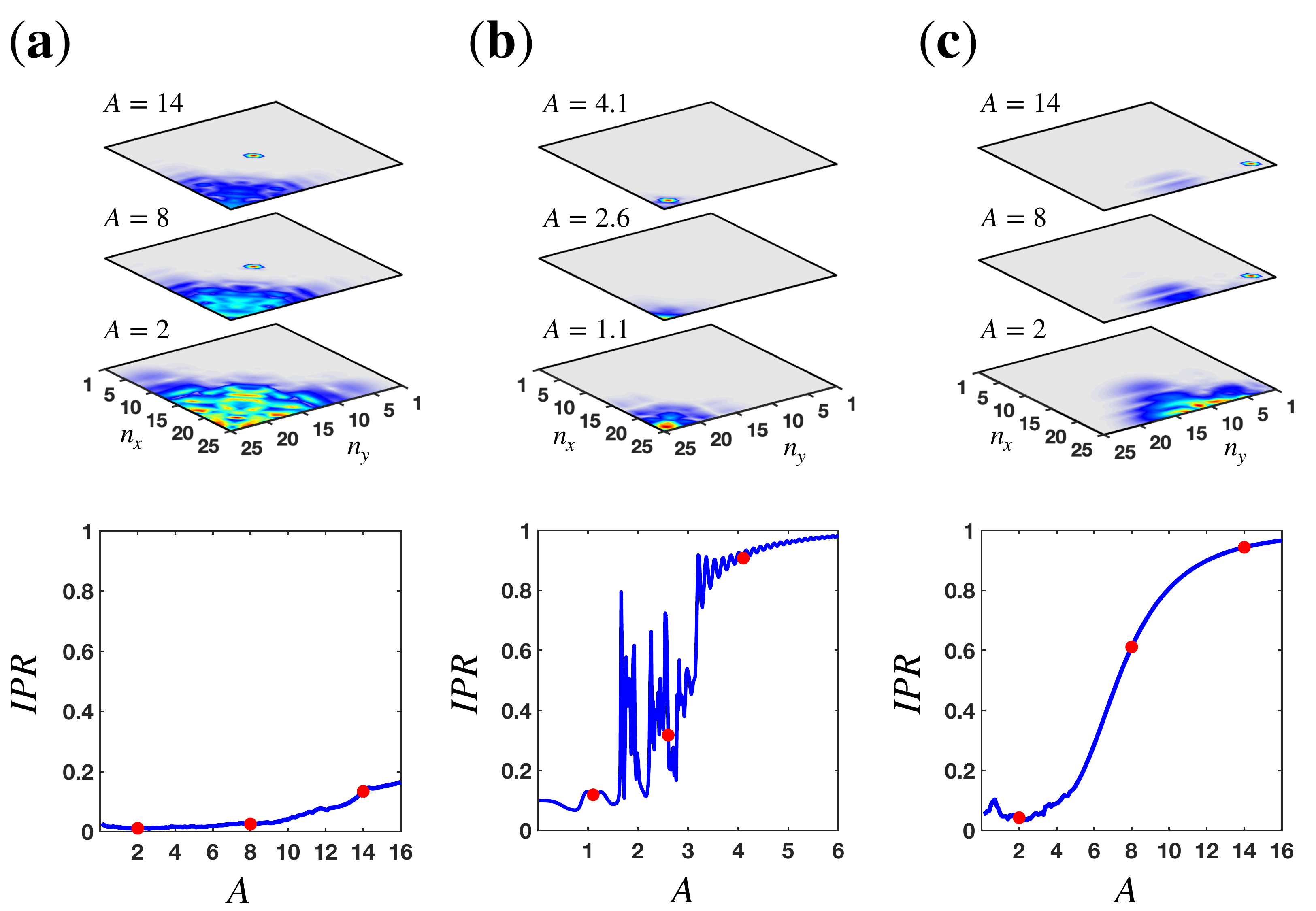}
    \caption{Comparison of the self-trapping tendency for different locations of excitation channel, under fixed $h=0.4$. Normalized amplitudes \( |\psi_{n_{x},n_{y}}(z=5)| \) for various values of $A$ (top row) under single-site excitation (a) at the center, $n_{x_0}=n_{y_0}=13$; (b) near corner (A), $n_{x_0}=n_{y_0}=23$; and (c) near corner (B), $n_{x_0}=23$ and $n_{y_0}=3$. $IPR$ for $z=5$ is shown as a function of \( A \) in bottom row for all cases. Red dots indicate the values of $IPR$ for the corresponding values of $A$ in top row.}
    \label{fig_5}
\end{figure}

First, we focus on the case of initial excitation at the center of the lattice, i.e., $  n_{x_{0}}=n_{y_{0}} = 13 $ for Hermitian and non-Hermitian lattices and our results are shown in Fig.~\ref{fig_4}. In particular, in Fig.~\ref{fig_4}(a) for the Hermitian case (\( h=0 \)), the wavefunction at a fixed propagation distance \( z = 5 \), \( |\psi_{n_{x},n_{y}}(z=5)| \), is extended for low values of the excitation amplitude \( A \). When \( A \) exceeds a particular value, the effect of Kerr nonlinearity becomes significant, leading to self-trapping of the wavefunction, almost entirely, at the position of initial excitation. The transition between delocalization and self-trapping can be quantified by the increase in the output (i.e., at \( z=5 \)) value of the inverse participation ratio ($IPR$). This metric is widely used to indicate localization and, for a given two-dimensional wavefunction \( \psi_{n_{x},n_{y}} \), is defined as:
\begin{equation}
        IPR \equiv \frac{\sum_{n_{x}=1}^{N}\sum_{n_{y}=1}^{N} |\psi_{n_{x},n_{y}}|^4}{\left(\sum_{n_{x}=1}^{N}\sum_{n_{y}=1}^{N} |\psi_{n_{x},n_{y}}|^2\right)^2},
\end{equation}
ranging from \( 1/N^{2} \) for a fully extended state \( \psi_{n_{x},n_{y}} \propto 1/N \), to 1 for a state localized to a single site, \( \psi_{n_{x},n_{y}} \propto \delta_{n_{x},m_{x}} \delta_{n_{y},m_{y}} \). Interestingly, this behavior changes dramatically when coupling asymmetry is introduced, through non-zero values of $h$. For the case of \( h=0.2 \), as shown in Fig.~\ref{fig_4}(b), it is evident that propagation predominates over self-trapping for a much higher range of excitation amplitudes \( A \), compared to the Hermitian lattice. 
Our results regarding higher values of the non-Hermiticity parameter $h$, and different locations of the single-channel excitation, are presented in Fig.~\ref{fig_5}. More specifically, for \( h=0.4 \), Kerr nonlinearity can no longer induce self-trapping up to $z=5$ for single-site excitation at the center of the lattice, as shown in Fig.~\ref{fig_5}(a). Instead, the NHSE dominates, forcing the wavefunction to delocalize toward the corner (A) of the lattice. This behavior persists even as the excitation amplitude \( A \) increases, within a reasonable range of physically relevant values, as indicated by the consistently low values of the $IPR(z=5)$. Nevertheless, as shown in Fig.~\ref{fig_5}(b), the competition between nonlinearity and coupling asymmetry leads to different propagation characteristics when the initial excitation is near the corner (A) of the lattice, i.e., $n_{x_0}=n_{y_{0}}=23$. Since the linear eigenmodes of the system are localized near this location, self-trapping to the initial excitation-site becomes possible above an amplitude threshold,  contrary to the case of Fig.~\ref{fig_5}(a).   {Interestingly, the pronounced peaks of $IPR(z=5)$ for $A\lesssim4$ correspond to cases, where, after the wavefunction gets delocalized from the site of initial excitation due to coupling asymmetry, it becomes trapped in lattice sites \textit{closer to the corner (A)} ($n_{x_0}=n_{y_0}=25$)  of the lattice, for finite propagation intervals.} At this point, it is interesting to also consider the case of an initial excitation near corner (B), i.e., at $n_{x_0}=23$ and $n_{y_0}=3$. There, delocalization of the wavefunction is promoted only along the spatial dimension $n_y$, since the wavefunction is already on the side of the lattice towards which the coupling along the $n_x$ direction is higher. As shown in Fig.~\ref{fig_5}(c), although self-trapping is possible, it requires considerably higher excitation amplitude. This represents an intermediate case between those discussed in Fig. ~\ref{fig_5}(a) and Fig. ~\ref{fig_5}(b).

To conclude, the aforementioned examples highlight the intricate interplay between Kerr nonlinearity, which induces self-trapping of an initially localized wavepacket, and non-Hermiticity, which promotes wavefunction delocalization. Up to this point, it is clear that the antagonism between those two effects depends on three different factors; the non-Hermiticity parameter $h$, the location of the initial excitation $(n_{x_0}, n_{y_0})$, and its amplitude $A$. A natural and physically relevant question arising from our discussion is how to quantify such amplitude thresholds required to induce self-trapping at a single lattice site, for a given non-Hermiticity parameter $h$.

\subsection*{Amplitude thresholds for self-trapping}
In order to investigate this systematically, we study the dynamics in a 2D NLHN lattice, for single-site excitation at each site $(n_{x_0}, n_{y_0})$, gradually increasing the excitation amplitude $A$. For each value of $A$, we compute the output $IPR(z=5)$ of the wavefunction, $|\psi_{n_x, n_y}(z=5)|$, similarly to the previous subsection. We define the threshold $A_{th}$ for self-trapping as the lowest value of $A$ for which $IPR (z=5)> 0.8$. Such a criterion is meaningful as it signifies near-perfect self-trapping at a single site, and although the threshold choice is arbitrary, it does not compromise the generality of our results since any alternative criterion would lead to similar conclusions. Furthermore, to ensure that the high value of $IPR(z=5)$ arises from self-trapping to the site of initial excitation rather than localization to other sites toward the corner (A) of the lattice, we impose an additional constraint: the shift of the mean position of the wavefunction at $z=5$, $r \equiv \sqrt{(\langle n_x \rangle - n_{x_0})^2 + (\langle n_y \rangle - n_{y_0})^2}$, must satisfy $r < 1$. For reference, we mention that in the Hermitian case ($h=0$) the two aforementioned constraints yield an amplitude threshold of $A = 3.15$.

At first, we follow the previously described procedure for a lattice with $h= 0.2$, and present our results in Fig.~6(a) as a comprehensive map. As evident from this figure, self-trapping occurs at relatively low excitation amplitudes, around $A_{th} \sim 3$, for lattice sites adjacent to corner (A). However, the threshold increases to approximately $A_{th} \sim 4$ for lattice sites near corners (B) and rises dramatically to $A_{th} \sim 9.5$ for the bulk of the lattice. {This dependence of $A_{th}$ on the excitation site stems from the position-dependent tendency for delocalization in the 2D HN lattice. Initial single-channel excitations at sites adjacent to corner (A), where couplings are stronger in both transverse directions, exhibit minimal delocalization even in the linear regime and therefore require the lowest amplitudes for self-trapping when nonlinearity is present. Excitations at sites near corner (B), where couplings are stronger along only one direction, exhibit important diffraction only along one direction and thus require higher thresholds, while excitations at bulk sites or near corner (C), which rapidly delocalize in both directions, require even larger amplitudes for self-trapping.} 

When the non-Hermiticity parameter is increased to $h= 0.4$ [Fig.~6(b)], the previously observed asymmetry in amplitude thresholds becomes significantly more profound. While sites near corner (A) still require an excitation amplitude of $A_{th} \sim 3$ for self-trapping, the threshold exceeds $A_{th} = 10$ for sites near corner (B) and, more interestingly, becomes impossible for the bulk of the lattice. For this latter observation, we examined excitation amplitudes up to $A = 20$. At this point we note, that when asymmetric couplings only occurs  along one transverse direction, i.e., $h_x=0.4$, $h_y=0$, then self-trapping becomes feasible for all lattice sites, as presented in the SM.

In conclusion, in this subsection we have examined how Kerr nonlinearity induces self-trapping in a single-site when the excitation amplitude \( A \) exceeds a specific threshold, depending on the location of the initially localized excitation and the degree of non-Hermiticity, in a 2D NLHN lattice.
\begin{figure}
    \centering
    \includegraphics[width=0.7\textwidth]{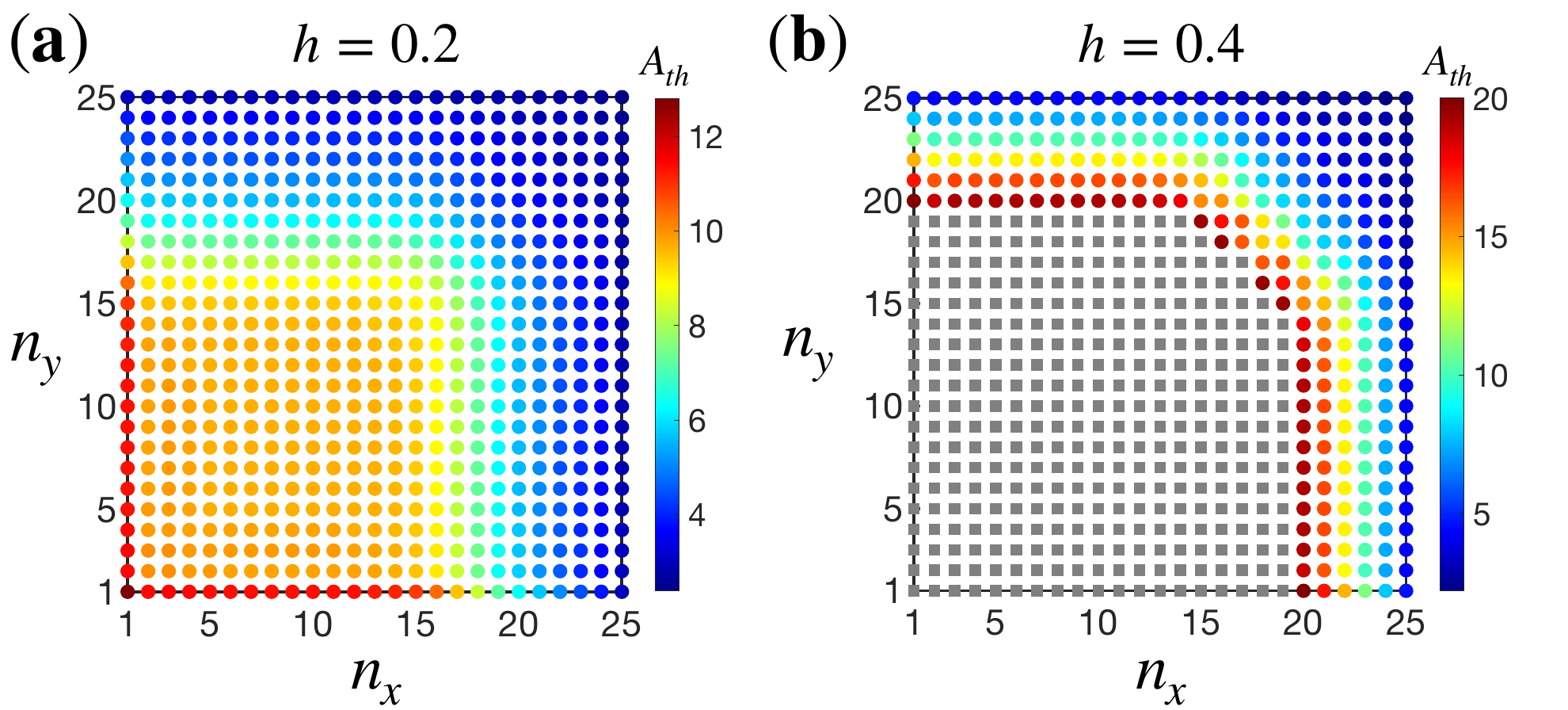}
    \caption{Amplitude thresholds $A_{th}$ required for single-site self-trapping, for all lattice sites (colormap). Two values of non-Hermiticity parameter are considered: (a) $h = 0.2$ and (b) $h = 0.4$. Gray squares in (b) indicate lattice sites where single-site self-trapping is impossible for excitation amplitudes up to $A = 20$.}
    \label{fig_6}
\end{figure}
\subsection*{Self-focusing and defocusing dynamics}
{
Up to this point, we have examined the propagation dynamics in 2D NLHN lattices specifically for an initially localized single-channel excitation, uncovering a rich competition between self-trapping and lattice's non-Hermiticity. As previously mentioned, for this type of excitation, the amplitude dynamics remain identical irrespective of the nature (focusing or defocusing) of the Kerr nonlinearity. Therefore, it is intriguing to investigate the dynamics under wider excitations to qualitatively explore the differences arising to the interplay of coupling asymmetry and Kerr nonlinearity, between the focusing ($g=1$) and defocusing ($g=-1$) cases.}

{Specifically, we consider an initial Gaussian excitation centered at the lattice site $(x_c, y_c)$:
\begin{equation}
    \label{Gauss}
    \psi_{n_x,n_y}(z=0)=A\frac{e^{-[(n_x - x_c)^2+(n_y - y_c)^2]}}{2 w^2},
\end{equation}
where $A$ is the amplitude and $w$ is the Gaussian width parameter. In Fig.~7, we present an illustrative comparison between the propagation dynamics in the focusing (top row) and defocusing (bottom row) Kerr-nonlinear regimes for a non-Hermiticity parameter $h=0.2$. The initial excitation parameters are set to $w=2$, $(x_c,y_c)=(13,13)$, with the amplitude $A$ chosen to yield a total optical power of $\mathcal{P}_{\text{HN}}(z=0)=20$ [Fig. 7(a)/(d)]. Additionally, in Fig.~8, we plot the evolution of the wavefunction’s mean positions, $\langle n_x \rangle=\langle n_y \rangle$, and their uncertainties, $\Delta n_x=\Delta n_y$, comparing the focusing and defocusing nonlinear-regime propagation dynamics, with the corresponding linear case.}

{In the focusing nonlinear case, the wavefunction initially undergoes self-focusing toward the lattice center, as shown in Fig.~7(b), also evidenced by the decrease in the position uncertainties [Fig.~8(b)] and directly observed. Subsequently, however, the wavefunction begins to spread [Fig.~7(c)], and the mean positions begin to rise, though remaining consistently lower than in the linear scenario [Fig.~8(a)]. Notably, after a certain propagation distance, the wavefunction separates into two distinct parts, similarly to what discussed in the first subsection for single-site excitations: one fraction remains localized near the initial excitation location, while another part propagates toward corner (A) of the lattice [Fig.~7(d)], as reflected in increased position uncertainties for $z\gtrsim 5$, compared to the linear case [Fig. 8(b)].}
\begin{figure}[H]
    \centering
    \includegraphics[width=1\textwidth]{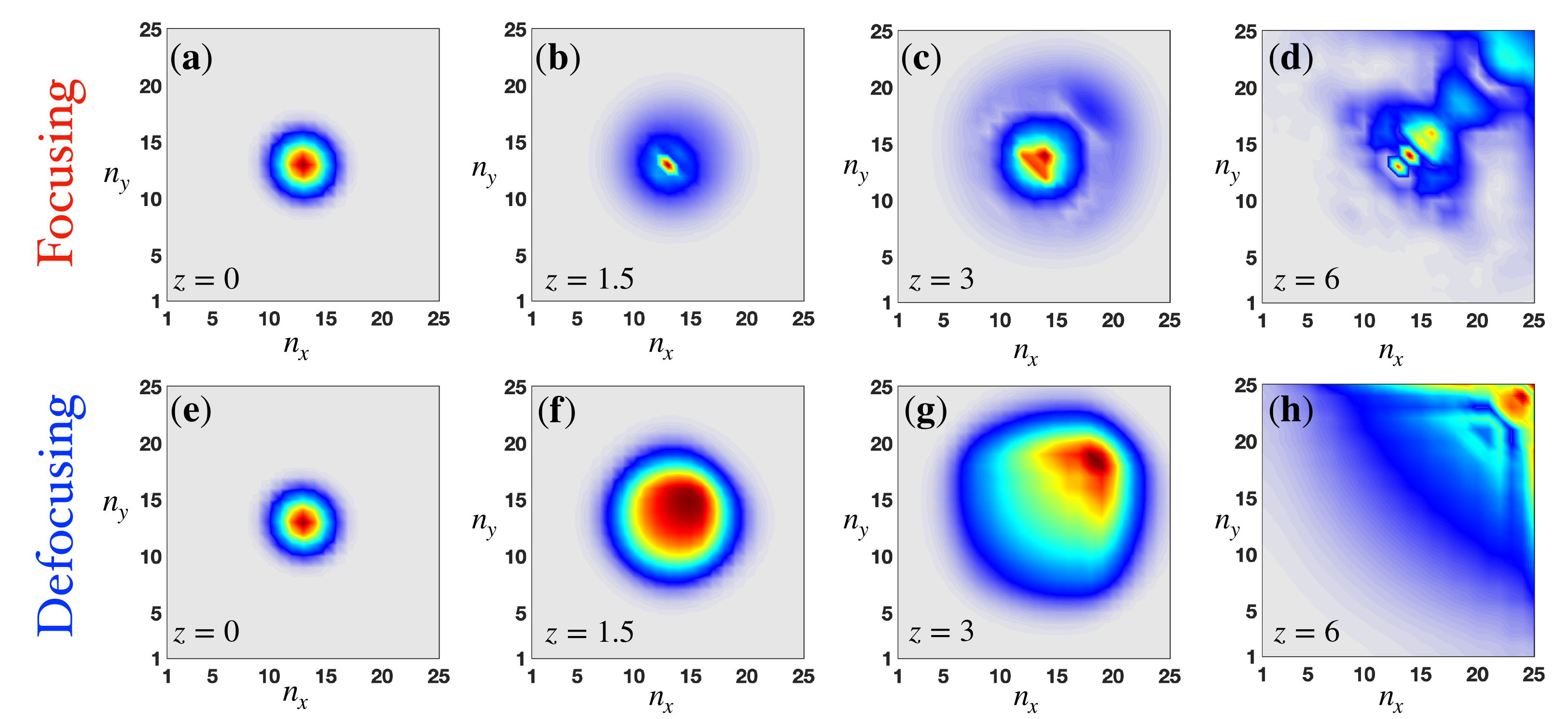}
    \caption{{Propagation dynamics in a 2D HN lattice for the focusing (top row) and defocusing Kerr-nonlinear  (bottom row) cases, with non-Hermiticity parameter $h = 0.2$ and initial condition \( \psi_{n_x,n_y}(z=0)=A{e^{-[(n_x - x_c)^2+(n_y - y_c)^2]}}/{2 w^2} \), with $x_c=y_c=13$, $w=2$  and $A$ such that $\mathcal{P}_{\text{HN}}(z=0)=20$. (a)–(d) Normalized complex amplitude $|\psi_{n_x, n_y}(z)|$ for the focusing case at propagation distances (a) $z = 0$, (b) $z = 1.5$, (c) $z = 3$ and (d) $z=6$. (e)-(h) Normalized complex amplitude $|\psi_{n_x, n_y}(z)|$ for the defocusing case at the same propagation distances.}}
    \label{fig_6}
\end{figure}
{In contrast to the single-site excitation scenario, the Gaussian initial condition [Eq.~\eqref{Gauss}] exhibits significantly different dynamics under defocusing Kerr nonlinearity. As illustrated in Figures 7 and 8, the wavefunction shifts away from its initial position markedly faster compared to the linear or focusing nonlinear cases, propagating preferentially toward corner (A) of the lattice, as reflected to increased mean positions. Importantly, it simultaneously exhibits stronger diffraction compared to the linear regime, as indicated by its higher position uncertainties. Thus, although the wavefunction mean position shifts toward the lattice's preferred corner, the presence of coupling asymmetry does not prevent a significant fraction of the wavefunction from spreading broadly across the lattice.}

{Here, we note that the dynamics remain qualitatively similar to what described above for Gaussian excitations centered at different $(x_c,y_c)$ across the lattice. It is also noteworthy that, as $w$ decreases, the differences between the focusing and defocusing cases become less pronounced, and for $w \lesssim 0.3$ the propagation dynamics of the two regimes are almost indistinguishable, converging to the single-site excitation scenario.}

\begin{figure}[H]
    \centering
    \includegraphics[width=0.63\textwidth]{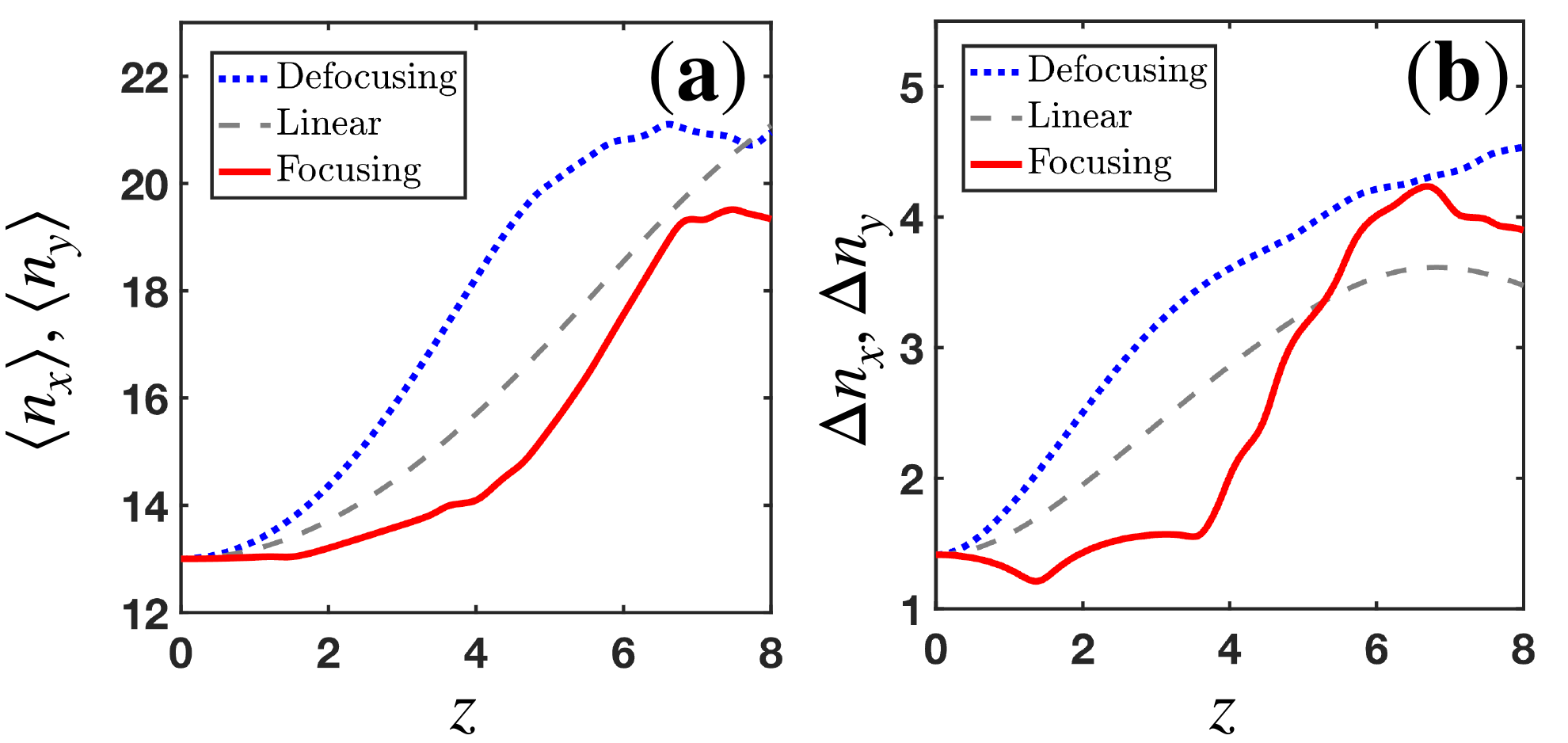}
    \caption{{Evolution of mean positions and uncertainties in a 2D HN lattice with non-Hermiticity parameter $h= 0.2$, for initial condition \( \psi_{n_x,n_y}(z=0)=A{e^{-[(n_x - x_c)^2+(n_y - y_c)^2]}}/{2 w^2} \), with $x_c=y_c=13$, $w=2$  and $A$ such that $\mathcal{P}_{\text{HN}}(z=0)=20$. (a) Mean positions \( \langle n_x \rangle \) and \( \langle n_y \rangle \)  and (b) position uncertainties \( \Delta n_x \) and \( \Delta n_y \) for the linear case (gray lines), focusing Kerr-nonlinear (red lines), and defocusing Kerr-nonlinear cases (blue lines).}}
    \label{fig_6}
\end{figure}
\subsection*{Two-dimensional skin solitons}
\begin{figure}[H]
    \centering
    \includegraphics[width=0.6\textwidth]{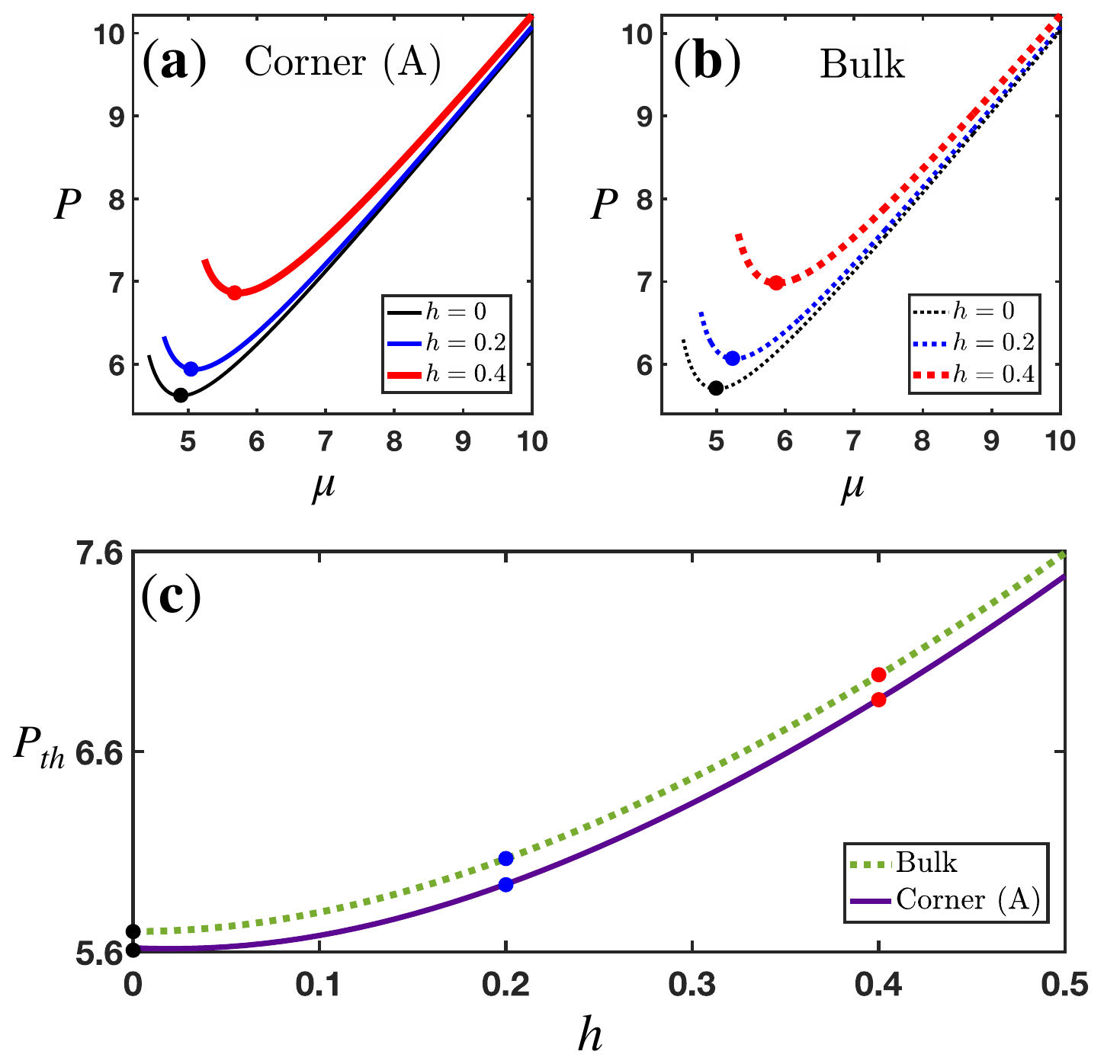}
    \caption{Power-eigenvalue $P-\mu$ diagrams for different families of solitons (a) near corner (A)  and (b) in the bulk, for a 2D NLHN lattice. Colored lines correspond to different values of the non-Hermiticity parameter $h$. (c) Power thresholds $P_{th}$ as a function of $h$, for solitons near corner (A) (blue line) and in the bulk (green line). Black, blue and red dots correspond to the dots of (a)-(b) indicating the value of $h$.}
    \label{fig_7}
\end{figure}

\begin{figure}[H]
    \centering
    \includegraphics[width=0.94\textwidth]{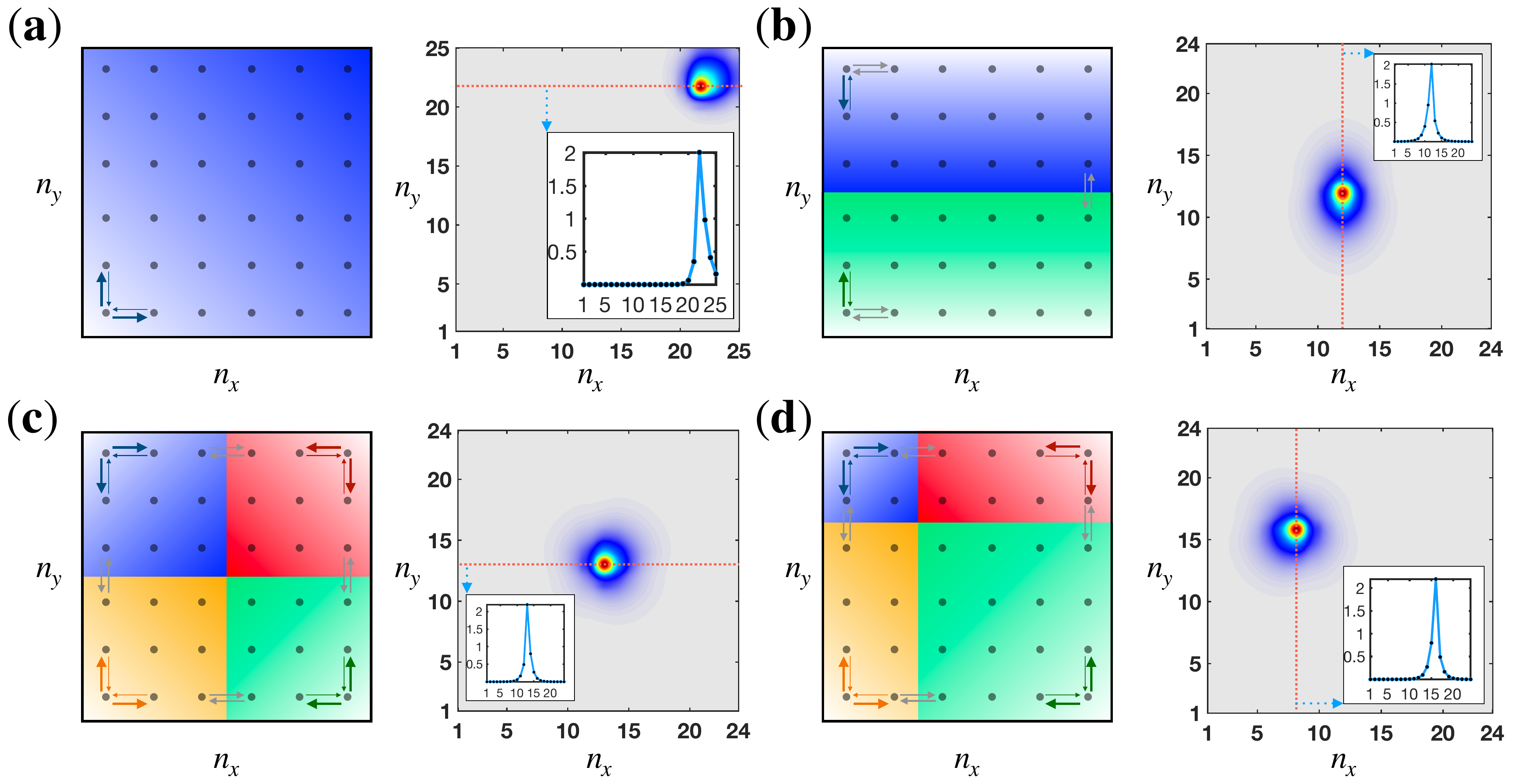}
    \caption{{Two-dimensional skin solitons in different lattice geometries. (a) 2D NLHN lattice with $N \times N = 25 \times 25$ sites and non-Hermiticity parameter $h=0.4$.. (b)–(d) Alternative lattices with asymmetric couplings: the directions of strong (weak) couplings are indicated by thick (thin) colored arrows in the left panels. At all boundaries between sublattices, the intra-sublattice couplings are reciprocal and set to $c=1$ (gray arrows). The right panels show representative soliton profiles $\big|\phi_{n_x,n_y}\big|$ for the corresponding geometries, calculated for lattices with $N \times N = 24 \times 24$ sites composed of 2D HN sublattices with $h=0.4$. In each right panel, an inset displays a cross-section along the red dashed line.} 
\label{fig_8}
}
\end{figure}

In previous subsections, we systematically discussed that Kerr nonlinearity induces self-trapping of the wavefunction to its initial position, for single-channel excitation, when the excitation amplitude \( A \) exceeds a threshold, which depends on the location of the initial excitation and the degree of coupling asymmetry.
This analysis naturally leads to the question of whether this lattice also supports soliton solutions, analogous to the skin solitons recently predicted theoretically~\cite{SL} and observed experimentally~\cite{SL_Exp} in 1D NLHN systems. 
In this section, we identify stationary soliton solutions within the 2D NLHN lattice, particularly for \textit{focusing} Kerr-nonlinearity ($g=1$). In particular, we consider solutions of Eq.~(\ref{kerr})  of the form $\psi_{n_x, n_y}(z) = \phi_{n_x, n_y} e^{i\mu z}$, where by $\phi_{n_x, n_y}$ and $\mu$ we denote the soliton profile and the soliton eigenvalue, respectively. Substituting this ansatz into Eq.~(\ref{kerr}) results in a system of nonlinear algebraic equations,

\begin{equation}
\begin{split}
e^h \phi_{n_x - 1 , n_y} + e^{-h} \phi_{n_x + 1 , n_y} 
+ e^h \phi_{n_x , n_y - 1} + e^{-h} \phi_{n_x , n_y + 1} 
+ |\phi_{n_x , n_y}|^2 \phi_{n_x , n_y} = \mu \phi_{n_x , n_y}.
\end{split}
\end{equation}
Using iterative numerical techniques~\cite{2D_Segev, Surface_Solitons, Solitons_Optics, Ziad}, we obtain lattice soliton solutions and thus we can relate their corresponding power ($P \equiv \sum_{n_x=1}^{N}\sum_{n_y=1}^{N}|\phi_{n_x,n_y}|^{2}$) to the soliton eigenvalue ($\mu$). Such power-eigenvalue diagrams are shown in Fig.~\ref{fig_7} for various values of $h$. Specifically, we analyze solitons positioned near corner (A) [Fig.~\ref{fig_7}(a)] and near the center of the lattice [Fig.~\ref{fig_7}(b)]. In both cases, it is evident that the solitons exhibit power thresholds. We have also identified soliton solutions located near corners (B) and (C) of the lattice, having power–eigenvalue diagrams very similar to the solitons located near the center of the lattice. For a given value of $h$, the power thresholds for solitons near corners (B) and (C) differ by less than $10^{-2}$ from those of the solitons near the center. However, the eigenvalues at these thresholds differ more, reflecting variations in the spatial distributions of the soliton profiles.
Additionally, for both soliton locations, we investigate the dependence of the power thresholds $P_{th}$ on the non-Hermiticity parameter $h$. As illustrated in Fig.~\ref{fig_7}(c), the power thresholds for solitons located near corner (A) and near the center of the lattice increase monotonically with $h$. Specifically, we confirm that this dependence is well approximated by a parabolic form, $P_{th}\approx ah^2+bh+c$. {In Fig.~\ref{fig_8}(a), we present a pertinent example of a soliton profile $|\phi_{n_x,n_y}|$ located near corner (A), for $h=0.4$. In this case, as well as for solitons at other locations in the lattice, the solutions exhibit spatial asymmetry toward corner (A) due to the asymmetric couplings. As expected, we further verified that the degree of asymmetry of the soliton solutions increases with $h$.} Also, we investigated the stability of the soliton solutions by computing the deviation parameter defined as $\delta\phi_{n_x,n_y}(z) \equiv \max\left(\big||\phi_{n_x,n_y}(z)| - |\phi_{n_x,n_y}(0)|\big|\right)$, at a large propagation distance $z = 25$. This parameter quantifies the deviation of the evolved soliton profile from its initial distribution. For a non-Hermiticity parameter $h=0.2$, solitons at all lattice locations remain stable, exhibiting very small deviations, with $\delta\phi \sim 10^{-7}$. However, at a higher non-Hermiticity value of $h=0.4$, soliton stability varies significantly with lattice position: solitons near corner (A) remain highly stable, with extremely low deviations ($\sim 10^{-10}$), while deviations increase progressively when moving from corner (B) ($\sim 10^{-7}$) toward the lattice center ($\sim 10^{-4}$), reaching their largest value near corner (C) ($\sim 10^{-3}$).

{Finally, we examined the formation of \textit{skin solitons} in other 2D lattices with asymmetric couplings under focusing Kerr nonlinearity, beyond the prototypical 2D NLHN lattice. As shown in Fig.~10, skin solitons also exist in alternative non-Hermitian lattices composed of properly engineered sublattices, where the global coupling asymmetry is directed, for example, toward a line, the lattice center, or other targeted regions~\cite{topo10}. These geometries are schematically illustrated in the left panels of Fig.~10(b)–(d). For each geometry, the right panels of Fig.~10(b)–(d) display a representative skin soliton localized near the region toward which the couplings are stronger, where the spatial asymmetry is clearly evident. For simplicity, all examples use 2D HN sublattices with non-Hermiticity parameter $h=0.4$. We numerically confirmed that skin solitons in these complex lattices also exhibit power thresholds. Consequently, the two-dimensional geometry offers a broad range of possibilities for supporting distinct asymmetric skin soliton solutions in complex non-Hermitian lattices, extending well beyond the 2D NLHN lattice~\cite{Li2025}.}

\section*{Discussion}

In this work, we have investigated the propagation dynamics of a generalized two-dimensional nonlinear Hatano–Nelson lattice, focusing on the intricate antagonism between the skin effect and self-trapping induced by Kerr nonlinearity. In the linear regime, wavefunctions are driven toward a lattice's preferred corner by the NHSE, whereas the presence of nonlinearity substantially suppresses this delocalization tendency. In particular, for single-site excitation, we have shown that the self-trapping behavior strongly depends on the amplitude of the initial excitation, its position, and the degree of coupling asymmetry. Specifically, excitations near the corner where the linear modes are localized exhibit self-trapping above relatively low amplitude thresholds. By contrast, in the lattice bulk, asymmetric couplings effectively oppose self-trapping, leading to delocalization even at higher excitation amplitudes; for very strong non-Hermiticity, self-trapping in these bulk regions becomes practically impossible. Furthermore, to highlight the differences in this antagonism between focusing and defocusing Kerr-nonlinear regimes, we examined the propagation dynamics of broader initial excitations, revealing a stronger tendency for delocalization in the defocusing case, associated with enhanced diffraction.

Additionally, we have identified and characterized two-dimensional stationary bright soliton solutions in various lattice locations by calculating their associated power–eigenvalue diagrams. These solutions, which extend the concept of skin solitons in two dimensions, consistently display spatial asymmetry biased toward the lattice corner favored by asymmetric couplings. Interestingly, our study revealed that the soliton power thresholds increase parabolically with the non-Hermiticity parameter $h$. Remarkably, we have shown that such skin soliton solutions also exist for a plethora of asymmetric lattices with more complicated geometries, apart from the Hatano–Nelson lattice. Consequently, our results set the stage for further exploration of skin soliton formation in non-Hermitian lattices~\cite{Li2025}. 

{Importantly, recent advances in non-Hermitian photonics have enabled the experimental realization of nonlinear non-Hermitian systems in synthetic-mesh lattices~\cite{2D_4, SL_Exp}. In this platform, Kerr nonlinearity has been implemented through an opto-electronic feedforward loop, enabling the observation of skin solitons in 1D HN lattices~\cite{SL_Exp}. Moreover, the same synthetic-mesh architecture has been employed to realize linear 2D non-Hermitian lattices, where both spatial dimensions are synthesized by mapping fast and slow time bins of coupled fiber loops onto a two-dimensional lattice~\cite{2D_4}. Given these developments, we believe that extending such schemes to implement Kerr nonlinear effects in 2D non-Hermitian lattices is within reach of current experimental capabilities. In conclusion, our results may provide useful insights into transport and soliton formation in higher-dimensional nonlinear non-Hermitian lattices, which are highly relevant to the emerging field of nonlinear non-Hermitian optics and photonics.}

\section*{Acknowledgements}

The authors acknowledge financial support from the European Research Council (ERC) through the Consolidator Grant Agreement No. 101045135 (Beyond Anderson). E.T.K. and I.K. further acknowledge funding from the Stavros Niarchos Foundation (SNF) and the Hellenic Foundation for Research and Innovation (H.F.R.I.) under the 5th Call of the “Science and Society” Action, titled “Always strive for excellence – Theodoros Papazoglou” (Project No. 11496, “PSEUDOTOPPOS”). Computations for this paper were partially conducted on the Metropolis cluster, supported by the Institute of Theoretical and Computational Physics, Department of Physics, University of Crete.

\section*{Author contributions}

E.T.K. and I.K. performed the numerical simulations and analyzed the results. K.G.M. conceived the idea and supervised the project. All authors contributed to writing the manuscript.

\end{document}